\documentclass[10pt]{IEEEtran}

\makeatletter
\def\ps@headings{%
	\def\@oddhead{\mbox{}\scriptsize\rightmark \hfil \thepage}%
	\def\@evenhead{\scriptsize\thepage \hfil \leftmark\mbox{}}%
	\def\@oddfoot{}%
	\def\@evenfoot{}}

\makeatother

\usepackage[english]{babel}
\usepackage{lipsum}
\usepackage{indentfirst}
\usepackage[colorlinks=false,linkcolor=black, citecolor=black]{hyperref}
\usepackage{algorithm}
\usepackage{mathtools}
\usepackage{balance}
\usepackage{commath}
\usepackage{amsmath}
\usepackage{algpseudocode}
\usepackage{float}
\usepackage{graphicx}
\usepackage{bm}
\usepackage[skip=2pt]{caption}
\usepackage{array}
\usepackage{verbatim}
\usepackage{xcolor}

\newcommand{\AAA}[1]{\textcolor{blue}{{\sf AA: #1}}}
\newcommand*{\affmark}[1][*]{\textsuperscript{#1}}

\begin{document}
	
	\title{RF Jamming Classification using Relative Speed Estimation in Vehicular Wireless Networks}
	\vspace{-5mm}
	\author{\IEEEauthorblockN{Dimitrios Kosmanos \affmark[1], Dimitrios Karagiannis \affmark[1], Antonios Argyriou \affmark[1], Spyros Lalis \affmark[1] and Leandros Maglaras \affmark[2]}\\
    
		{$^{\dag}${\affmark[1]Department of Electrical and Computer Engineering, University of Thessaly, Greece}}\\
        {$^{\dag}${\affmark[2]Department of Computing Technology, De Montfort University, Leicester, UK}}\\

		\vspace{-5mm}
	}

\maketitle

\markboth{ Latest version \today}{IEEE TVT, Latest version \today}

\begin{abstract}
Wireless communications are vulnerable against radio frequency (RF) jamming which might be caused either intentionally or unintentionally. A particular subset of wireless networks, vehicular ad-hoc networks (VANET) which incorporate a series of safety-critical applications, may be a potential target of RF jamming with detrimental safety effects. To ensure secure communication and defend it against this type of attacks, an accurate detection scheme must be adopted. 

In this paper we introduce a detection scheme that is based on supervised learning. The machine-learning algorithms, K-Nearest Neighbors (KNN) and Random Forests (RF), utilize a series of features among which is the metric of the variations of relative speed (VRS) between the jammer and the receiver that is passively estimated from the combined value of the useful and the jamming signal at the receiver. To the best of our knowledge, this metric has never been utilized before in a machine-learning detection scheme in the literature. Through offline training and the proposed KNN-VRS, RF-VRS classification algorithms, we are able to efficiently detect various cases of Denial of Service Attacks (DoS) jamming attacks, differentiate them from cases of interference as well as foresee a potential danger successfully and act accordingly. 

\end{abstract}


\section{INTRODUCTION}
\vspace{1ex}
Autonomous vehicles capable of navigating unpredictable real-world environments with little human feedback are a reality today~\cite{iov2017}. Autonomous vehicle control imposes very strict requirements on the security of the wireless communication channels ~\cite{guardian} used by a fleet of vehicles~\cite{vehicle-platooning}. Moreover, with the Intelligent Vehicle Grid technology, the vehicle becomes a formidable sensor platform, absorbing information from the environment or from other vehicles ( called as Internet of Vehicles IoV) and feeding it to other vehicles and infrastructure so as to assist in safe navigation and traffic management.
As a result, the term {\it smart city} has been coined to describe the city of tomorrow in which modern intelligent technologies, such as
IT communication systems, sensors, machine learning, data analytics, come together to provide better services to the citizens.
\par
Wireless communications, however, are vulnerable against a wide range of attacks.
An attack that is particularly hard to detect in every wireless network is the RF jamming attack~\cite{hamieh2009detection}. In a VANET, RF attack detection is even more challenging due to the constant and rapid changes in topology and the high mobility of the nodes as well as due to the presence of a variety of different jammers
. These jamming scenarios affect either the communication between vehicles (V2V communication) or the communication between the vehicles and the roadside units, namely RSUs (V2R communication).

Over the last few years, there have been several experimental approaches for jamming detection~\cite{azogu2013new},~\cite{hamieh2009detection},
~\cite{punal2014machine},~\cite{xu2005feasibility}, some of which suggest the use of machine learning~\cite{SybilAttacks2011},~\cite{punal2014machine}. However, only \cite{punal2014machine} examines closely the adoption of machine learning techniques for jamming detection. Furthermore, none of the above works that propose machine-learning based schemes, have investigated the relative speed, which is an application layer metric, because vehicular wireless channels exhibit rapid changes. However with our work, we prove that this metric can be used in a realistic scenario with a minimum number of assumptions leading to an increase in the accuracy of the classification procedure.

In this paper we propose the use of RF signals for estimating the relative speed between the jammer and the receiver and use the  variations of it as an extra feature in the classification process.
The proposed metric is combined with the physical layer metrics leading to a cross-layer approach for offline training. This set of cross-layer features are utilized for the classification of different jamming scenarios. In the general case, jamming reduces the receiver signal to interference and noise ratio (SINR), a problem that can be addressed with classic communication algorithms. In several applications, however, it is critical to detect accurately the presence of a jammer ($J_x$), i.e. the precise reason behind the reduction in the signal-to-noise ratio (SNR) and the packet-delivery-ratio (PDR), and even more so, the nature of the attack. So, it is difficult to determine whether the reason for the SINR reduction is a jamming attack or unintentional interference. This is the main motivation for the relative speed metric input in a jamming detection scheme. However, our results indicate that the proposed scheme can effectively  differentiate a case of jamming attack from that of an interfering wireless source. This is a very crucial point because each of the two cases can be treated with a different solution. 
Especially, in the case of interference, an Interference Cancellation (IC) model~\cite{ici2016} or a spectral evasion (channel surfing, spatial retreats) scheme could be applied. 
\subsection{Contributions}
Our contribution in this paper is two-fold:
\begin{itemize}
\item A novel detection scheme is introduced that leverages the use of a new metric from the application layer that is passively estimated by the physical layer, namely \textit{the variations of relative speed (VRS)}. 
\item Based on a series of cross-layer data (among which is the VRS metric that is calculated using the estimated relative speed) detection of a potential threat as well as differentiation between a case of jamming and a case of interference is achieved.
 \end{itemize}
Our jamming detection scheme could be applicable to a platoon of vehicles in which an exterior or an interior attacker can cause significant instability in the  Cooperative Adaptive Cruise Control (CACC) of the vehicle stream ~\cite{platoons2015}. 
For the validation of our approach, one interference-only scenario and two jamming attack scenarios have been created and tested.
\par
The rest of this paper is structured as follows: Section II provides an overview of the related work in the domain of attack (not only jamming) detection, Section III describes  the adopted topology and the channel model, Section IV describes the methodology behind the estimation of the relative speed, Section V presents the proposed machine-learning based jamming detection system, Section VI describes the simulation setup, Section VII presents the experimental results and comparisons and Section VIII summarizes the significance of our approach and concludes our work. 

\section{RELATED WORK}
\vspace{1ex}
\parskip 0pt
Machine-learning based approaches for attack detection in vehicular ad-hoc networks have been primarily reported in~\cite{punal2014machine} and~\cite{SybilAttacks2011}.
Pu{\~n}al et al.~\cite{punal2014machine} use several metrics like the Noise and Channel Busy Ratio (CBR), Packet Delivery Ratio (PDR), Maximum Inactive Time (Max IT), Received Signal Strength (RSS) to detect attacks with machine learning  techniques and examine the cases of reactive and constant jammers. 
Azogu et al.~\cite{azogu2013new} have proposed a new mechanism, called the Hideaway Strategy according to which all nodes should remain silent while the network is under a jamming attack. Bi{\ss}meyer et al.~\cite{bissmeyer-2010-intrusion} propose a detection scheme that is based on the verification of vehicle movement data and on the notion that a certain space will be occupied by only one vehicle at a certain time. Hamieh et al.~\cite{hamieh2009detection} focus on the detection of the reactive jammer. The proposed method, is based on the correlation coefficient (CC) and the error probability (EP). Each node compares the calculated value of CC with the EP and the network is considered under a jamming attack if CC\textgreater EP.
Malebary et al.~\cite{malebary2016jamming} propose a two-phase jamming detection method that utilizes metrics such as the RSS, the Packet Delivery/Send Ratio (PDSR) and the Packet Loss Ratio (PLR) as well consistency checks to distinguish a jamming from a no-jamming situation. 
%
\par
RoselinMary et al.~\cite{roselinmary2013early} present an approach that is based on the detection of malicious and irrelevant packets using the number of broadcast packets per second (frequency) and the velocity of the vehicle that the packets are sent from. 
%
Shafiq et al.~\cite{shafiq2005detection} investigated an attack detection approach based on the number of packets that are received from each vehicle, thus indicating an attack if this number is greater than the threshold value. 
Xu et al.~\cite{xu2005feasibility} indicated the inability of the PDR to differentiate jamming from interference cases. For that reason, two detection schemes are proposed. The first one utilizes signal strength measurements as a consistency check to determine if the PDR value is due to jamming or interference. The second uses location information as the consistency check. Several jamming attack models are presented and evaluated.
\par
Sharanya et al.~\cite{sharanya2006classifying} propose the use of the Support Vector Machine (SVM) algorithm with Modified Fading Memory (MFM) so as to classify legitimate and malicious nodes. The purpose of the MFM is to reduce the computational overhead for the machine learning algorithm by only considering as eligible nodes those in the range of the VANET communication only for limited time.\par
Last, Karagiannis et al.~\cite{Karagiannis-elsevier} propose a RF jamming attack detection scheme using an unsupervised learning with clustering. The novelty of this paper is that the relative speed metric is utilized between the jammer and the receiver, along with other parameters, in order to differentiate intentional from unintentional jamming as well as identify the unique characteristics of each jamming attack. However, this relative speed metric is obtained from the on-board wireless communication devices at the receiver vehicle and is not estimated through the wireless medium.\par
We must mention here that in all the previous work that propose machine-learning based schemes the relative speed has not been considered as a classification feature. Our proposed system is the first in literature that uses the point-to-point RF communication in order to estimate the relative speed metric. Variations of relative speed are used for effectively differentiating interference from jamming, by distinguishing the unique characteristics of each attack. Moreover, we use a supervised learning scheme, in which the input data is utilized by the wireless communication between transmitter and receiver with the presence of a jammer. This supervised learning scheme feeds that data back into the supervised learning algorithms as training data and uses these algorithms to accurately classify all the kinds of attacks.

\section{SYSTEM MODEL}
\label{model}
\subsection{Topology}
\begin{figure}[t]
\centering
\includegraphics*[keepaspectratio, width=1.0\linewidth]{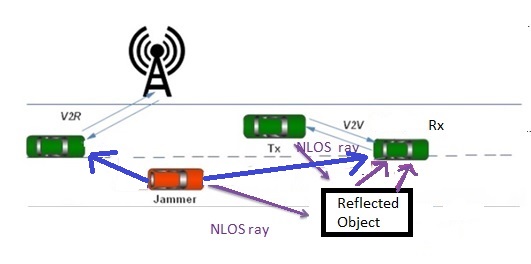}
\caption{Topology. Blue arrows represent the LOS V2V or V2R wireless communication. While, purple arrows represent the NLOS components that are caused by a Reflected Static Object.}
\label{fig:topology}
\end{figure}

The topology we adopt in our work (Figure~\ref{fig:topology}) involves a moving vehicle, namely $R_x$, that serves as the target of the jammer, another vehicle or a RSU (namely $T_x$) that is used as the transmitter of the useful signal and the jamming vehicle that tries to intervene in the communication between $R_x$ and $T_x$. The travelling speed of the $R_x$, namely $u_{R_x}$, is equal with the travelling speed of the $T_x$, namely $u_{T_x}$, and is bound to the limitations of an urban environment. Last, there is a static object in the area that causes the multipath fading effects from reflection. Upon spotting its target, the jammer begins following it and starts jamming either continuously or periodically (in order to stay undetected for as long as possible). This jamming behavior summarizes all the potential Denial of Service Attack (DoS) jamming attacks. 
\subsection{Rician Fading Model}
\label{model: B}
In our work, we adopt the Rician fading model that is a channel model which includes path loss and also Rayleigh fading ~\cite{tse}. When a signal is transmitted, whether it is a useful signal or a jamming one, this channel adds multipath fading in addition to thermal noise. It is assumed that a LOS ray and $N-1$ NLOS rays exist in the area.
The baseband signal that the receiver receives from the jammer is 
\begin{equation}
\begin{aligned}
\label{eq:model_B}
y(t)&=\sum_{n=0}^{N-1}((h_1(t,n) x_{pilot}[N-n] \sqrt{P_1} \\
&+ h_2(t,n)s[N-n]  \sqrt{P_2}) + w(t)
\end{aligned}
\end{equation}
where
\small
\begin{equation}
\begin{aligned}
\label{eq:h1}
h_1(t,n)&=Ray_1(n) \\ 
&+ \frac{1}{dist_1^2(t)}) e^{j(2\pi/\lambda ) (f_c+ f_{d,max} \cos{\theta_{1}}) \tau_{1}(n) } \delta{(t-\tau_{1}(n))}
\end{aligned}
\end{equation}
and
\begin{equation}
\begin{aligned}
\label{eq:h2}
h_2(t,n)&=Ray_2(n) \\
&+ \frac{1}{dist_2^2(t)}) e^{j(2\pi/\lambda ) (f_c+ f_{d,max} \cos{\theta_{2}}) \tau_{2}(n)} \delta{(t-\tau_{2}(n))}
\end{aligned}
\end{equation}
\normalsize
Where, \textit{$h_1(t,n),h_2(t,n)$} are the Rician fading channel models between transmitter-receiver and jammer-receiver respectively. This type of channel model includes path loss and also Rayleigh
fading. Moreover, \textit{$Ray_1(n)$}, \textit{$Ray_2(n)$} are complex Gaussian variables capturing the Rayleigh fading between transmitter - receiver and jammer - receiver, and $x_{pilot}[N-n],s[N-n]$ are the symbols that are transmitted from the transmitter and the jammer respectively, for which the BPSK modulation is assumed. This modulation scheme is preferred because achieves lower bit error rate and we want a reliable communication between $T_x$ and $R_x$. Moreover, this modulation scheme is the most robust in a high interference environment. In the above equation, $f_c$ is the carrier frequency, $f_{d,max}$ is the maximum Doppler shift. \textit{$P_1$} and \textit{$P_2$} are the transmission power per symbol of the useful and of the jamming signal respectively and \textit{w(t)} is the channel noise at the time instant \textit{$t$}. The terms $d_s,d_j$ correspond to the distance between the transmitter and the reflected object and between the jammer and the reflected object.
While the terms $r_{1n}(t),r_{2n}(t)$ correspond to the distance between the transmitter and the receiver and between the jammer and the receiver. In \eqref{eq:h1}, \eqref{eq:h2} the distance that travels the LOS rays is represented with the $dist_1(t)=r_{1n}(t), dist_2(t)=r_{2n}(t)$ respectively. On the other hand, the distance that the NLOS rays travel is represented with the $dist_1(t)=2d_{s}-r_{1n}(t), dist_2(t)=2d_{j}-r_{2n}(t)$  respectively. Moreover, $f_c$ is the carrier frequency, $f_{d,max}$ is the maximum Doppler shift, $\theta_{1}$ is the incidence AOD between the vector of speed $\vec{u}_{T_x}$ and the vector of signal sent from the transmitter, $\theta_{2}$ is the incidence AOD between the vector of speed $\vec{u}_{J_x}$ and the vector of signal sent from the jammer, $(\tau_{1}= dist_1(t)/c,\tau_{2}= dist_2(t)/c)$ is the excess delay time for the transmitter and jammer signal ray (that may be caused due to ground reflection) and $t$ is the current time instant. For the rest of the paper, we will use the parameter $\gamma_{1}$\footnote{$(\gamma_{1}=(Ray_1(n) + \frac{1}{r_{1n}^2(t)}))$} as the transmitter - receiver complex amplitude associated with the LOS path and the parameter $\gamma_{2}$\footnote{$(\gamma_{2}=(Ray_2(n) + \frac{1}{r_{2n}^2(t)}))$} as the jammer - receiver complex amplitude. The above complex amplitude values are known at the receiver.

\subsection{System Overview}
In our system model, a fixed number of known pilot symbols 
are being sent through the wireless IEEE 802.11p standard ~\cite{malebary2016jamming} over consecutive time instants from the transmitter to the receiver. At the same time, the jammer simultaneously transmits over consecutive time instants random jamming symbols to the receiver. 
Using the pilots, the LOS channel and the $N-1$ NLOS channels between $J_x - R_x$ will be estimated by the receiver. 

\par

\textbf{System Description}: The basic idea of our system is to first estimate the relative speed between the jammer and the receiver, exploiting the RF Doppler shift. Subsequently, we use the variations of the estimated relative speed as a new feature in a supervised machine learning algorithm for RF jamming attack detection. Along with the relative speed from the application layer, we use cross-layer data that we obtain from the physical layer, such as the Received Signal Strength and Interference (RSSI), the Signal to Interference and Noise Ratio (SINR) and the Packet Delivery Ratio (PDR). Two classification algorithms are investigated, namely the k-Nearest Neighbors (KNN) and the Random Forests (RF) algorithm respectively.

\subsection{Jamming Scenarios}
 We assume that the jammer continuously transmits with
the appropriate transmission power, with the purpose of overloading the
wireless medium creating, thus a DoS attack \cite{ddos-attack}.
\par 
In our work, we also created three different attack scenarios -namely \textbf{Interference Scenario}, \textbf{Smart Attack Scenario} and \textbf{Constant Attack Scenario} - each representing a jamming attack case that could affect a VANET in real life. 
\label{scenarios}
\label{scenario: 1}

In the Interference Scenario, we assume that a moving, intentional jammer is not present in the network so as to evaluate the efficiency of our method in differentiating jamming from the interference level of the external environment. The vehicle travels, when, at some point, passes through an area with significant RF interference by which its communication with the other vehicles or with the RSU is affected.

\label{scenario: 2}
In the Smart Attack Scenario, the performance of a more intelligent jammer is evaluated~\cite{kosmanos2016mimo}. Specifically a smart jammer starts following the victim-vehicle, while transmitting a jamming signal. When the jammer reaches its target at a distance of about $10m$, retreats to a safe position and transmits in a reactive way. The most common alteration in literature is when the jammer keeps changing its transmission power, thus achieving the same disrupt or thwart in the communication (DoS attack) without the need of changing its distance from the target. With our Smart Attack, we aim at affecting the communication of the $T_x$-$R_x$ pair, with the jammer detection being more difficult, while using the relative speed metric as feature. For that reason the "smart" jammer retreats to a safe position. The jammer is designed to start transmitting upon sensing energy above a certain threshold. We set the latter to $-86 dBm$ as it is empirically determined to be a good tradeoff between jammer sensitivity and false transmission detection rate. If the detected energy exceeds the threshold during a certain time span $(T_{detection}= 12\mu s)$, an ongoing 802.11p transmission is assumed by the jammer and starts its transmission for a duration of $(T_{duration}= 84\mu s)$.
This smart jammer is designed to affect the header of the 802.11p frame sent from $T_x$ to $R_x$.

\label{scenario: 3}
In the Constant Attack Scenario, we study the case of a constant jammer that follows the victim-vehicle while transmitting constantly at a minimum power. When the jammer reaches its target, begins transmitting constantly with its full power without any intention to stay undetected as at the Smart Attack Scenario.

\section{ Relative Speed Metric Estimation}
\label{RS-ESTIMATION}
In this section, we present the relative speed metric $(\Delta{u})$ that indicates the relative speed between the jammer and the victim's vehicle. Based on the obtained values, the Variations of Relative Speed (VRS) metric is created and then used for the classification. To the best of our knowledge, this metric has never been used before for jamming detection. The relative speed metric as firstly defined in ~\cite{Karagiannis-elsevier} is:
\begin{equation}
\Delta{u}=|\vec{u}_{J_x}-\vec{u}_{R_x}|
\end{equation}
Where, $\vec{u}_{J_x}$, $\vec{u}_{R_x}$ the speed of jammer and the speed of receiver respectively. \par
From the equation ~\eqref{eq:h2}, the $N$ multipath combined channels $(h_1+h_2)$ can be estimated, using a MMSE estimator ~\cite{tse}. After, exploiting the Doppler phenomenon for modeling the LOS $h_{2}^{LOS}$ channel between jammer and receiver, we can estimate the defined above relative speed metric.
\section{PROPOSED DETECTION SYSTEM BASED ON SUPERVISED LEARNING}
\subsection{Proposed Algorithm}
\vspace{1ex}
\parskip 0pt
\hypersetup{linkcolor=red}
\par
To make our detection method robust, apart from the physical and network metrics used in related works, we introduce and use the VRS metric from the application layer that can be effectively estimated by the RF signals interchange in the physical layer (see Section \ref{RS-ESTIMATION}). Our method uses this new metric, as an extra feature in a cross-layer approach, along with other metrics from the physical layer for the classification process. All these metrics are presented in Table ~\ref{array:metrics}.
\begin{center}
\begin{table}

\centering
\caption{Metrics that are jointly processed by the classification algorithms}
\label{array:metrics}
 \begin{tabular}{|p{1.8cm}|c|} 
 \hline
  \textbf{   Metrics }  \\ [0.5 ex]
\hline\hline
  $\Delta{u}$ \\ 
 \hline
    $RSSI$    \\
 \hline
    $SINR$  \\
 \hline
    $PDR$   \\
 \hline
 \end{tabular}
\end{table}
\end{center}
\par
To create the VRS metric for the classification process we, initially, make three fundamental assumptions:
\begin{itemize}
\label{basic_assumptions}
  \item When the relative speed is equal to zero and remains unchanged, it indicates the existence of a constant jammer that follows the victim-vehicle.
  \item When the relative speed is not equal to zero and remains unchanged, it indicates the absence of a moving jammer as the relative speed is equal to the speed of the vehicle.
  \item When the relative speed is not equal to zero for a period of time and then becomes zero while remaining unchanged, it indicates the existence of a jammer that begins following the target after reaching it.
\end{itemize}
The common characteristic of these assumptions is that the speed of the participating non-malicious vehicles remains unchanged and is always greater than zero.
\par
However, in a real-life scenario, such as the one that we study, the speed - and as a consequence the relative speed - may not remain constant during the observation period. In other words, if we want to accurately model an urban environment, we have to consider the fact that the vehicles can alter their traveling speed. 
To handle these real-life situations, while still using the previously presented assumptions, we introduce the \textbf{Variations of Relative Speed Algorithm} \ref{alg:the_alg} (VRS Algorithm).
\par
The \textit{VRS Algorithm} detects changes in the relative speed of the provided observations. To ensure that, the relative speed and the speed from the previous as well as the subsequent observations are used along with a series of control flow statements.
The algorithm is divided into two main parts, with the first examining the case in which the relative speed value is not equal to zero and the second examining the opposite case, each one with its own logical checks to determine the existence of a threat or not.
\par
The $\Delta{u}$ and $u$ variables represent an array of estimated relative speed values and travelling speed values respectively, \textit{M} is the number of the available observations upon which the algorithm operates, \textit{vrs} is an array used to store the classification result (A for attack or NA for not attack) of the current observation and the \textit{trigger} is a binary variable which indicates the presence of a jammer (value is equal to 1) or its absence (value is equal to 0). The \textit{NA} and \textit{A} values are two extreme and distinct values able to differentiate the attack from the no attack cases and guide the classification process. 

\subsection{Supervised Learning Algorithms}
The supervised learning methods that are used are KNN~\cite{sutton2012introduction} and Random Forests~\cite{liaw2002classification}. Their choice does not affect the efficiency of our algorithm as our proposed feature is not constrained by the type of the supervised learning algorithm that is used. The VRS Algorithm~\ref{alg:the_alg} generates the new metric which is used as an extra feature in the classification process without being affected at all by the machine learning algorithms that are utilized. Both supervised learning techniques are very popular, with the KNN being robust against noisy training data like the ones obtained from a real-life urban environment and Random Forests being one of the most accurate algorithms, due to the fact that it reduces the chance of over-fitting (by averaging several trees, there is a significantly lower chance of over-fitting). As it is previously stated, our detection scheme is currently based on offline training that leverages the use of a dataset of collected measurements in order to train the classifier.

\begin{algorithm}[!htb]
\caption{VRS Algorithm}
\label{alg:the_alg}
\begin{algorithmic}[1]
\State $M = $ number of observations 
\State $vrs = matrix(nrow = M, ncol = 1)$
\State $k = 1$
\If{$\textit{$\Delta{u}$[k] == $\Delta{u}$[k+1]}$}
	\State $vrs \gets NA$
	\State $trigger = 0$ 
\ElsIf{$\textit{$\Delta{u}$[k] $ \neq$ $\Delta{u}$[k+1]}$}
	\State $vrs \gets A$
	\State $trigger = 1$ 
\EndIf
\State
\State $k++$
\While{$(k < M) $}
    \If{$\textit{$\Delta{u}$[k] $\neq$ }0$}
    	\If {$\textit{$\Delta{u}$[k] $\neq$ $\Delta{u}$[k-1]}$}
        	\If {$\textit{$\Delta{u}$[k] == $u$[k]}$}
        		\State $vrs \gets NA$
        		\State $trigger = 0$
     		\ElsIf{$\textit{$\Delta{u}$[k] $\neq$ $u$[k]}$}
    			\State $vrs \gets A$
				\State $trigger = 1$ 
			\EndIf
      \ElsIf{$\textit{$\Delta{u}$[k]==$\Delta{u}$[k-1]}$}
    		\If {$\textit{$\Delta{u}$[k] $\neq$ $u$[k]}$}
    			\State $vrs \gets A$
				\State $trigger = 1$
			\ElsIf{$\textit{$\Delta{u}$[k] == $u$[k]}$}
			        \If{$\textit{hasNext == T}$}
			\If{($\textit{$\Delta{u}$[k-1]==$u$[k-1]} $\&\&$ $
			\\ 
			 \hspace{10mm} $\textit{$\Delta{u}$[k+1]==$u$[k+1]}$)}
				\State $vrs \gets NA$
        		\State $trigger = 0$
        		\Else
        			\State $vrs \gets A$
					\State $trigger = 1$
				\EndIf
				\ElsIf{$\textit{hasNext == F}$}
			\If{$\textit{trigger == 0}$}
				\State $vrs \gets NA$
        		\State $trigger = 0$
        	\Else
        		\State $vrs \gets A$
				\State $trigger = 1$
			\EndIf
		\EndIf
	\EndIf
\EndIf
\algstore{myalg}
\end{algorithmic}
\end{algorithm}

\begin{algorithm}[!htb]
  \ContinuedFloat
  \caption{VRS Algorithm (continued)}
  \begin{algorithmic}
      \algrestore{myalg}
\ElsIf{$\textit{$\Delta{u}$[k] == 0}$}
\If{$\textit{$u$[k] $\neq$ 0}$}
		\State $vrs \gets A$
		\State $trigger = 1$
\ElsIf{$\textit{$u$[k] == 0}$}
		\If{$\textit{$\Delta{u}$[k-1] == $u$[k-1]}}$
			\If{$\textit{trigger == 0}$}
				\State $vrs \gets NA$
        		\State $trigger = 0$
        	\Else
        		\State $vrs \gets A$
				\State $trigger = 1$
			\EndIf
		\ElsIf{$\textit{$\Delta{u}$[k-1] $\neq$ $u$[k-1] }$}
			\State $vrs \gets A$
			\State $trigger = 1$
		\EndIf
	\EndIf
\EndIf
\EndWhile
\State
\Return $\textit{vrs}$

\end{algorithmic}
\end{algorithm}

\section{Simulation Setup}
\label{assumptions}

\begin{figure}[!htb]
\centering
\includegraphics[width = 8cm, height = 5cm]{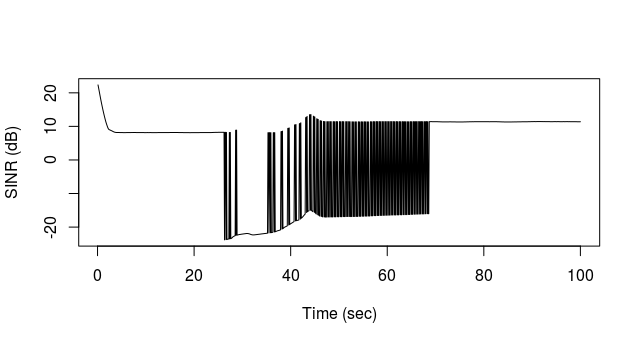}
\caption{SINR vs Time for the Rician Fading Model in the Interference Scenario}
\label{fig:modelB_1}
\end{figure}
\begin{figure}[!htb]
\centering
\includegraphics[width = 8cm, height = 5cm]{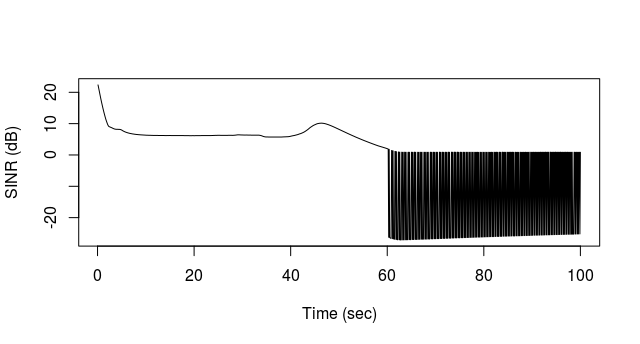}
\caption{SINR vs Time for the Rician Fading Model in the Smart Attack Scenario}
\label{fig:modelB_2}
\end{figure}
\begin{figure}[!htb]
\centering
\includegraphics[width = 8cm, height = 5cm]{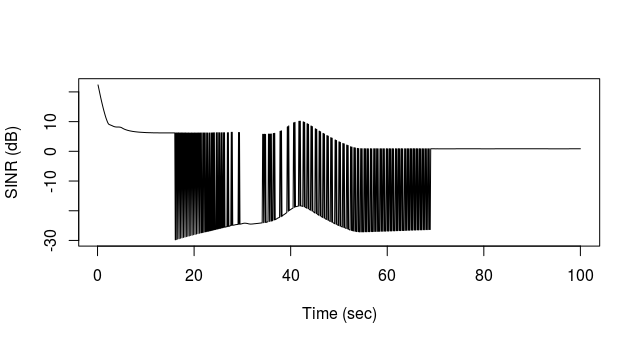}
\caption{SINR vs Time for the Rician Fading Model in the Constant Attack Scenario}
\label{fig:modelB_3}
\end{figure}

Figures~\ref{fig:modelB_1} - ~\ref{fig:modelB_3}, illustrate the behavior of the jammer by presenting the SINR versus Time for every one of the three scenarios namely \textbf{Interference Scenario}, \textbf{Smart Attack Scenario} and \textbf{Constant Attack Scenario} .

\subsection{Supervised Learning Testing Cases}
Beyond the scenarios that we use to evaluate the performance of the overall system, we also created a series of test cases that allow us to explore deeper the proposed system depending on the set of observations that is utilized for both training and testing. 

These cases only affect how the training and testing is performed, without any further implications in the scenarios. They are created in such a way so as to provide us with a comparison between the use or not of the VRS metric in the classification process under various circumstances, that would, in turn, highlight its significance.

\begin{itemize}
\label{cases}
  \item \textit{Train and test with data from the same speed value with the VRS metric \textbf{(Same\_KNN-VRS and Same\_RF-VRS case)}:} the prediction model is trained and tested using observations collected under the speed of 15 m/s, with the use of the VRS metric. To avoid testing with "previously seen data", thus leading to biased classification results, we have to ensure that the training and testing sets are completely separated. 
  \item \textit{Train and test with data from the same speed value without the VRS metric \textbf{(Same\_KNN and Same\_RF case)}:} similar to the previous case, with the only difference being the omission of the VRS metric in the classification process.
  \item \textit{Train and test under different speed values with the VRS metric \textbf{(Different\_KNN-VRS and Same\_RF-VRS case)}:} the previously trained prediction model is tested using data that was collected under a speed of 25 m/s, that is under a speed different from the one the training of the prediction model was based on.
  \item \textit{Train and test with data from different speed values without the VRS metric \textbf{(Different\_KNN and Different\_RF case)}:} similar to the previous case but without the utilization of the VRS metric as an extra feature in the classification process.
  \item \textit{Train and test with normalized data from the same speed value with the VRS metric \textbf{(Norm\_KNN-VRS and Norm\_RF-VRS case)}:} the data is normalized prior to its use training and in testing. By normalization, we refer to the process of changing the data so as to belong in the 0 - 1 range. Both training and testing are conducted on data collected under a speed of 15 m/s but without common observations in the two sets (as stated before).
  \item \textit{Train and test with normalized data from the same speed value without the VRS metric \textbf{(Norm\_KNN and Norm\_RF case)}:} similar to the previous case but without the VRS metric.
\end{itemize}

\subsection{Detection System Assumptions}
\label{sim_assumptions}
Regarding the details of our simulation setup, the \textit{speed} of the vehicles involved in the legitimate communication $(u_{T_{x},R_{x}})$, the initial \textit{distance between the jammer and the pair of $R_x$ - $T_x$} $(dist_{initial})$, the \textit{distance that separates the receiver from the transmitter} throughout the course of the simulation $(dist_{T_{x},R_{x}})$ as well as the power of all the the transmitted signals $(P_{T_{x},J_{x}})$ and the reference distance $(dist_{ref})$ with which the path loss component is estimated, are presented in Table~\ref{array:res}.
\par
The power of all the transmitted signals is measured in milliwatts (mW) and is converted in the dBm scale prior to using it in the algorithm. The signal that is transmitted from both the jammer and the transmitter consists of streams that are 500 bits long. For each one of the three scenarios, a number of 1000 packets is transmitted. Using a time sample of 0.1 sec, we simulate the system for 100 seconds (for each scenario) and obtain 1000 measurements (for each scenario). 
\par
We used the Simulation of Urban Mobility (SUMO) and the OMNET++/Veins ~\cite{veins}.
SUMO is adopted as our traffic simulator and OMNET++ is used to simulate the wireless communication. Part of the Erlangen city map is used for conducting the simulations. The evaluation parameters in the Veins simulator are also presented in Table~\ref{array:res}.
\begin{center}
\begin{table}

\centering
\caption{Simulation Parameters}
\label{array:res}
 \begin{tabular}{||c c||} 
 \hline
 \textbf{Evaluation Parameters in Veins Simulator}& \textbf{Values}  \\ [0.5 ex]
\hline\hline
 $u_{T_{x},R_{x}}$ & 15m/sec. \\ 
 \hline
 $dist_{T_{x},R_{x}}$ & 35m  \\
 \hline
 $dist_{initial}$ & 200m \\
 \hline
 $P_{T_{x},J_{x}}$ & 100mW  \\
 \hline
 Minimum sensitivity $(P_{th})$ & -69dBm to -85dBm  \\
 \hline
  Transmission Range & 130-300 meters \\
  \hline
  $f_c$ & 5.9GHz \\
  \hline
  Doppler shift for $\Delta{u}=120km/h$ & $\pm 655.5$ Hz  \\
  \hline
 $ dist_{ref}$ & 100m \\
  \hline
 \end{tabular}
\end{table}
\end{center}

\section{SIMULATION RESULTS}
\label{results}
To underline the significance of our proposed system, we proceed and compare the cases presented previously. In particular, for each supervised learning testing case presented in section~\ref{cases}, we execute a simulation which lasts 300 seconds and is equally split in the three jamming scenarios discussed in Section~\ref{scenarios}, so that the first 100 sec represent the Smart Attack Scenario, the next 100 sec the Interference Scenario and the last 100 sec the Constant Attack Scenario. All the above Scenarios are independent each other and simply presented at consecutive time instants.
\par
Prior to presenting the classification results, we have to define the size of the training and testing sets as well as the total number of observations used, so as to make them more interpretable. Each simulation, that is each case from~\ref{cases}, utilizes a set of 3000 observations, equally split into the three attack scenarios examined. To avoid overfitting\footnote{Overfitting occurs when the classifier tends to memorize the training set and thus generalize poorly when facing previously unseen data}, only 30\% of the total number of the observations is used for training while the remaining 70\% for testing.
\par
Based on the ratio above, in our simulations, the number of the observations in the training set is 941 (that is 293 observations from the Interference Scenario, 319 from the Smart Attack Scenario and 329 from the Constant Attack Scenario) whereas the number of the observations in the testing set is 2059 (that is 703 observations from the Interference Scenario, 685 from the Smart Attack Scenario and 671 from the Constant Attack Scenario), randomly chosen but almost equally split among the three scenarios in both cases. 

To present the classification results, the \textit{confusion matrix} is used~\cite{kohavi1998confusion}. Each row of the matrix represents the instances belonging to a predicted class while each column represents the instances in an actual class. To evaluate the performance of our detection system in the various scenarios previously described, we will use the \textit{accuracy of the prediction model}. Accuracy is a measure that is obtained from the confusion matrix and \textit{is equal to the ratio of all the correctly predicted labels over all the predictions.} The correctly predicted labels are the labels of the main diagonal of the confusion matrix.

\subsection{Same\_KNN/RF-VRS and Same\_KNN/RF cases}
\label{case1vs2}
Starting from the first case, the accuracy of the prediction model achieved is equal to 82.27\% for the KNN and 80.04\% for the Random Forests algorithm. An example of the above defined confusion matrix for the calculation of the accuracy of our prediction model is the subsequent confusion matrix for the KNN.

\begin{center}
\begin{tabular}{ |p{1.8cm}|c|c|c|c| } 
\hline
Scenario& \multicolumn{1}{p{1.5cm}|}{Interference}  & \multicolumn{1}{p{1.5cm}|}{Smart Attack} &\multicolumn{1}{p{1.5cm}|}{Constant Attack}\\
  \hline
     Interference& 703 & 0 & 0 \\ 
   \hline
    Smart Attack& 0 & 494 & 174 \\
  \hline
    Constant Attack& 0 & 191 & 497 \\  
  \hline
\end{tabular}
\captionof{table}{Confusion matrix for the Same\_KNN-VRS case} 
\label{confusion_same_knn}
\end{center}

\par
On the contrary, when omitting the VRS metric, we not only observe a drop in the classification accuracy but also a high confusion between interference and jamming cases. The accuracy of the prediction model is, now, equal to 79.16\% and 76.54\% for the KNN and the Random Forests algorithms respectively. So, the impact of the VRS metric is evident. Apart from the fact that it increases the success rate of the classification (compared to the cases where the VRS metric is omitted) it \textit{ensures}, almost perfectly, the differentiation between the cases of intentional and unintentional jamming.
\par

\subsection{Different\_KNN/RF-VRS and Different\_KNN/RF cases}
\label{case3vs6}
As stated previously, these cases examine the situation in which training and testing are based on observations that were collected under different speed. The accuracy achieved while using the VRS metric as an extra feature in the classification process is equal to 66.97\% for KNN and 69.84\% for Random Forests respectively.
\par
\begin{figure}[!htb]
\vspace{-1em}
\centering
\includegraphics[width = 9cm, height = 5cm]{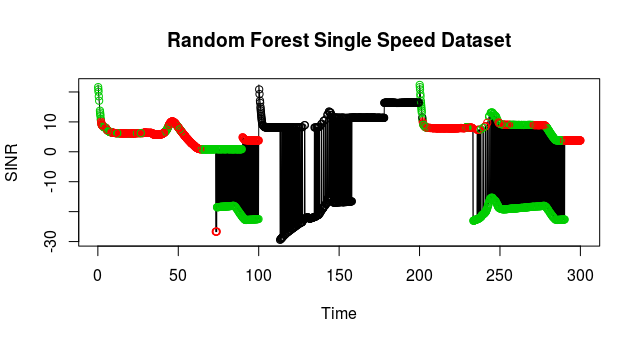}
\caption{SINR vs Time for the Different\_RF-VRS case, with the Smart Attack Scenario represented by the red, the Interference Scenario by the black and the Constant Attack Scenario by the green color.}
\label{fig:fig_3}
\end{figure}
\begin{figure}[!htb]
\vspace{-1.5em}
\centering
\includegraphics[width = 9cm, height = 5cm]{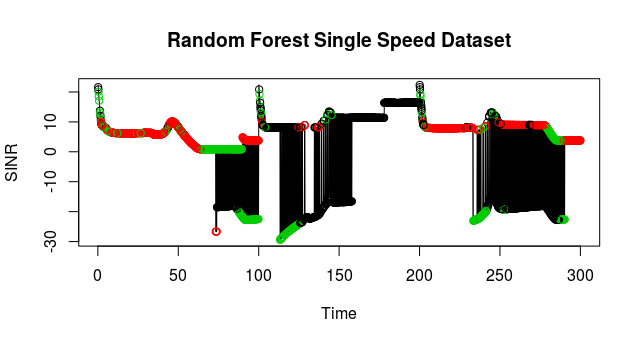}
\caption{SINR vs Time for the Different\_RF case, with the Smart Attack Scenario represented by the red, the Interference Scenario by the black and the Constant Attack Scenario by the green color.}
\label{fig:fig_4}
\end{figure}

On the other hand, when the VRS metric is not used, the accuracy of the prediction model is reduced to 56\% for the KNN and to 55.37\% for the Random Forests algorithm.
Figures~\ref{fig:fig_3} and~\ref{fig:fig_4} present the results for the Random Forests-based classification model.

\par
The color of the figures indicates the class in which each observation is predicted to belong to. \textit{The Smart Attack Scenario is represented by the \textit{red}, the Interference Scenario by the \textit{black} and the Constant Attack Scenario by the \textit{green} color.} The appearance of more than one colors in each scenario, that is in each 100 seconds (as described in~\ref{sim_assumptions}), indicates the existance of misclassifications.
\par
Based on the classification results presented above, we can reach an important conclusion. When testing the prediction model with observations from a different speed - compared to the one used in training - we observe an overall reduction in accuracy. With the VRS metric, not only the accuracy of the prediction model is significantly increased (in both supervised algorithms examined in this paper), but there is also a clear seperation between interference and jamming. \par
In addition to that, if we compare the previous classification results of the Same\_KNN and Same\_RF cases with that are derived when no normalization is applied to the data prior to their use, we observe that there is no significant increase in accuracy results. Thus we conclude to that a normalization of the measurements is not necessary.

\subsection{Same\_KNN/RF-VRS and Same\_KNN/RF cases for higher speed}
As it is already stated, our RF jamming attack detection system is based on offline training, using a dataset of measurements collected under a speed of 15m/s so as to train the classifier prior to its use for testing. For the sake of completion and in order to determine the behavior of our detection scheme when the training is conducted with data collected under a higher speed, we examine the Same\_KNN/RF-VRS and Same\_KNN/RF cases presented previously using as training data measurements from the 25m/s speed range.

For the Same\_KNN/RF-VRS case, the accuracy of the prediction model achieved is equal to 94.46\% for the KNN and 94.61\% for the Random Forests algorithm. For the Same\_KNN/RF case, on the other hand, the calculated accuracy is equal to 88.68\% for the KNN and 89.22\% for the Random Forests algorithm respectively. 
\par
From the classification results presented above two observations could be made. Our first observation could be that both the omission and the use of the VRS metric as an extra feature in the classification process lead to increase in accuracy compared to the results obtained in~\ref{case1vs2}. The use, however, of the VRS metric apart from the high classification accuracy, also leads to a clear differentiation between cases of intentional and unintentional jamming, something that is not obvious when the metric is omitted.
\par
\begin{figure}[!htb]
\vspace{-1.5em}
\centering
\includegraphics[width = 9cm, height = 5cm]{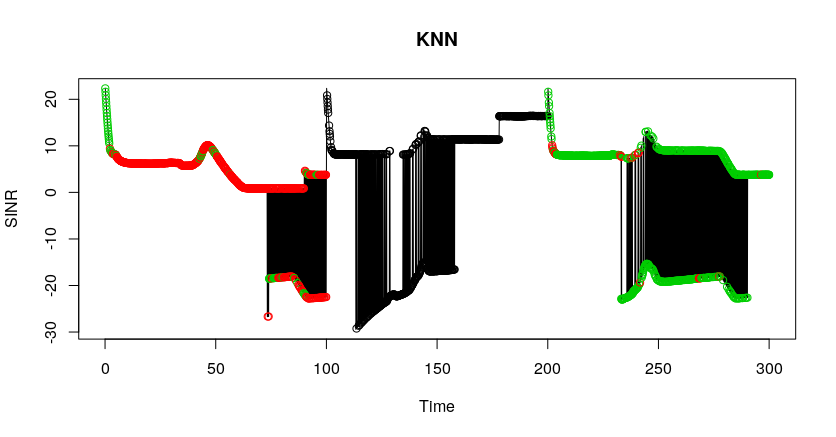}
\caption{Plot with train 25m/s for the Same\_KNN case, with the Smart Attack Scenario represented by the red, the Interference Scenario by the black and the Constant Attack Scenario by the green color.}
\label{fig:fig_6}
\end{figure}

Our second observation concerns the overall increase in classification accuracy when the training is done using data from a higher speed. The higher classification accuracy derives from the fact that the increase in speed adversely affects the effect of the jamming. For instance, in the Constant Attack Scenario the jammer overtakes the pair of $R_x$ - $T_x$ faster, in the Interference Scenario the the pair of $R_x$ - $T_x$ remains in the jamming area for a shorter period of time and in the Smart Attack Scenario the jammer reaches its target faster, thus the gradual effect of the jamming observed at lower speeds is greatly reduced. All the above lead to a significant increase in the quality of the measurements obtained, hence leading to higher classification accuracy as well as a to better distinction between the different types of jammers affecting the communication, as can be seen in Figure ~\ref{fig:fig_6} for the KNN algorithm.

In the following Table ~\ref{table:questions} we summarize the classification accuracy, exploiting the usage of the proposed VRS metric as an extra feature, achieved while training with measurements from a speed of 15m/s and a speed of 25m/s respectively.

\begin{center}
\begin{tabular}{ |p{2cm}|c|c|c| } 
\hline
 &\multicolumn{1}{p{1.6cm}|}{\textbf{Train with 15m/s}}  & \multicolumn{1}{p{1.6cm}|}{\textbf{Train with 25m/s}} \\
  \hline
     \textbf{Test with 15m/s} \textbf{(KNN)}& 82.27\% & 74.31\% \\ 
   \hline
    \textbf{Test with 15m/s} \textbf{(RF)}& 80.04\% & 74.41\% \\
  \hline
    \textbf{Test with 25m/s} \textbf{(KNN)}& 66.97\% & 94.46\% \\  
  \hline
    \textbf{Test with 25m/s} \textbf{(RF)}& 69.84\% & 94.61\% \\  
  \hline
\end{tabular}
\captionof{table}{Classification accuracy percentages while training lower and higher speed measurements respectively}
\label{table:questions}
\end{center}

\subsection{Comparison with Related Work}
\begin{figure}[!htb]
\vspace{-1em}
\centering
\includegraphics[width = 9cm, height = 5cm]{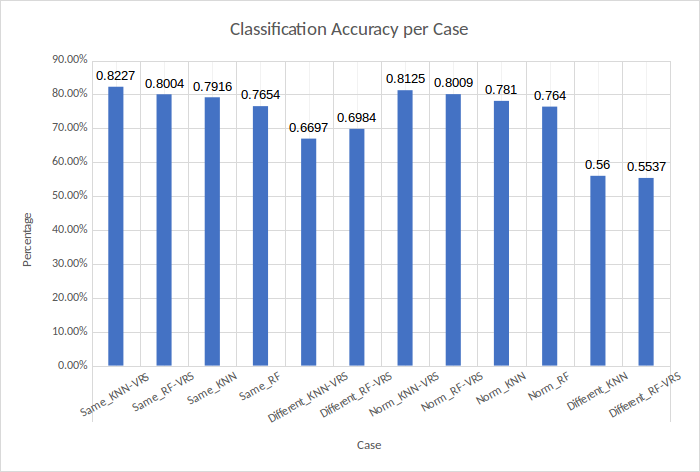}
\caption{Comparison between the standard KNN and RF classification algorithms and the proposed KNN-VRS, RF-VRS algorithms based on the accuracy percentage achieved in every case presented in all the above confusion matrices.}
\label{fig:fig_7}
\end{figure}

Figure~\ref{fig:fig_7} summarizes the classification accuracy percentages that are presented above. These are achieved by both the KNN and the Random Forests algorithm when based only on the features previously used in the literature for jamming attack detection~\cite{punal2014machine}, compared to the proposed approaches KNN-VRS and RF-VRS that use the VRS metric. The VRS metric increases the accuracy of the classifier and ensures almost perfect differentiation between cases of intentional and unintentional jamming. When using the VRS metric while testing with data from the same speed there is an increase up to about $4\%$ in the classification accuracy. When testing with data from a different speed, the increase in accuracy is even greater, up to about $14\%$. 

\section{CONCLUSIONS}
\vspace{-1ex}
In this paper, we presented a method for detecting a specific type of DoS attack, namely RF jamming, based on cross-layer supervised machine learning and by exploiting a novel metric from the application layer, the variations of the relative speed between the jammer and the target. The relative speed is passively estimated from the combined value of the desired and the jamming signal at the target vehicle combined with metrics from the network and physical layer. To evaluate the significance of the proposed metric and its estimation algorithm, we implemented three different scenarios - two with a jammer present and one with interference only.

With our work, we introduced a proactive approach against potential RF jamming attacks which is able to differentiate benign from malicious RF jamming, that is interference from jamming. Additionally, it is able to distinguish the unique characteristics of each attack, especially when the off-line training is conducted with a higher speed than $15m/s$.
Through our evaluation results, we were able to highlight the vital role of the relative speed and its variations from the application layer, in addition to other metric from the physical layer, in jamming detection and unintentional jamming cases differentiation, as well as in the overall increase in the prediction accuracy. 
\par
As part of our future work, we plan to use this classification process for a vehicular network with a large number of communicating nodes, as in a broadcast form. The target of this classification process will be the characterization of the behaviour of a node as malicious or not, mainly using the proposed VRS metric. The classification results can be collected from a Trusted Central Authority (TCA) in an area with V2X communication.

\balance
\bibliographystyle{IEEEtran}
\bibliography{bibfile}

\end{document}